\documentclass[dvips,6pt,cmp,aps,twocolumn,nofootinbib]{revtex4}

\usepackage[caption=false]{subfig}
\usepackage{amsmath,amssymb,amsfonts,epsf,epsfig,amsthm,bm,graphicx}

\begin{document}

\title{On the stationarity of linearly forced turbulence in finite domains}%

\author{E. Gravanis}
\email{elias.gravanis@cut.ac.cy}
\author{E. Akylas}
\email{evangelos.akylas@cut.ac.cy}
\affiliation{Department of Civil Engineering and Geomatics, Cyprus
University of Technology, P.O. Box 50329, 3603, Limassol, Cyprus}

\date{\today}%

\begin{abstract}
A simple scheme of forcing turbulence away from decay was introduced
by Lundgren some time ago, the `linear forcing', which amounts to a
force term linear in the velocity field with a constant coefficient.
The evolution of linearly forced turbulence towards a stationary
final state, as indicated by direct numerical simulations (DNS), is
examined from a theoretical point of view based on symmetry
arguments. In order to follow closely the DNS the flow is assumed to
live in a cubic domain with periodic boundary conditions. The
simplicity of the linear forcing scheme allows one to re-write the
problem as one of decaying turbulence with a decreasing viscosity.
Scaling symmetry considerations suggest that the system evolves to a
stationary state, evolution that may be understood as the gradual
breaking of a larger approximate symmetry to a smaller exact
symmetry. The same arguments show that the finiteness of the domain
is intimately related to the evolution of the system to a stationary
state at late times, as well as the consistency of this state with a
high degree of isotropy imposed by the symmetries of the domain
itself. The fluctuations observed in the DNS for all quantities in
the stationary state can be associated with deviations from
isotropy. Indeed, self-preserving isotropic turbulence models are
used to study evolution from a direct dynamical point of view,
emphasizing the naturalness of the Taylor microscale as a
self-similarity scale in this system. In this context the stationary
state emerges as a stable fixed point. Self-preservation seems to be
the reason behind a noted similarity of the third order structure
function between the linearly forced and freely decaying turbulence,
where again the finiteness of the domain plays an significant role.

\end{abstract}



\maketitle

\section{Introduction}

Maintaining a turbulent flow in a more or less stationary state, for
better statistics in experiment or convenience in theoretical
considerations, requires forcing the flow, that is feeding it energy
which balances dissipation happening at the smallest scales. In
numerical simulations of incompressible isotropic turbulent flows
one usually solves the Navier-Stokes equations in a cubic box (with
periodic boundary conditions). For an account of Direct Numerical
Simulation (DNS) methods see \cite{moin-mahesh}; for a recent review
on the current isotropic turbulence statistics from DNS see
\cite{DNS-review}. In most cases forcing takes the form of a force
term in wave number space (spectral space) which vanishes for all
but the smaller wave numbers i.e.\, one feeds energy at the largest
scales of the turbulent flow in the box. The general concept is that
the details of the larger scales are model dependent but the details
of all other scales, that is those where some universal laws may
hold, depend only on the intrinsic dynamics of the Navier-Stokes at
least for high Reynolds numbers. Presumably, by forcing turbulence
one achieves satisfactory results for given a resolution for higher
Reynolds numbers than in the freely decaying turbulence.

There have been developed various kinds of forcing schemes. The
simpler ones fiddle in a suitable manner the magnitude of velocity
field, or the total energy of the lower wave number modes, imitating
an energy input in the larger
scales~\cite{Siggia-forcing,she-forcing,chasnov-forcing,sullivan-forcing}.
These models can be regarded as essentially deterministic in the
sense that that there is no additional randomness introduced in the
problem. There are also deterministic models which explicitly
introduce a force term in the Navier-Stokes, whose details are
either postulated or derived by a postulated auxiliary model
\cite{kerr-forcing,siggia2-forcing,jimenez-forcing,overholt-forcing}.
In stochastic forcing
models~\cite{yahkot-forcing,eswaran-pope,alvelius} the details of
the force term are determined by additional random variables
following prescribed stochastic processes. Each of those models
suffers from one or more from a set of problems such as, excessive
fluctuations around stationarity, relatively long relaxation period
to stationarity, persistent anisotropy, excessive distortion of
large scale motions, introduction of irrelevant features in the
description of turbulence. A useful comparative discussion between
certain deterministic and stochastic models can be found in
\cite{TurbulenceIII}.

\begin{figure}[h]\vspace{-.0in}%
\begin{center}
  \subfloat[]{\includegraphics[height=0.265\textwidth,  angle=0,%
  clip=]{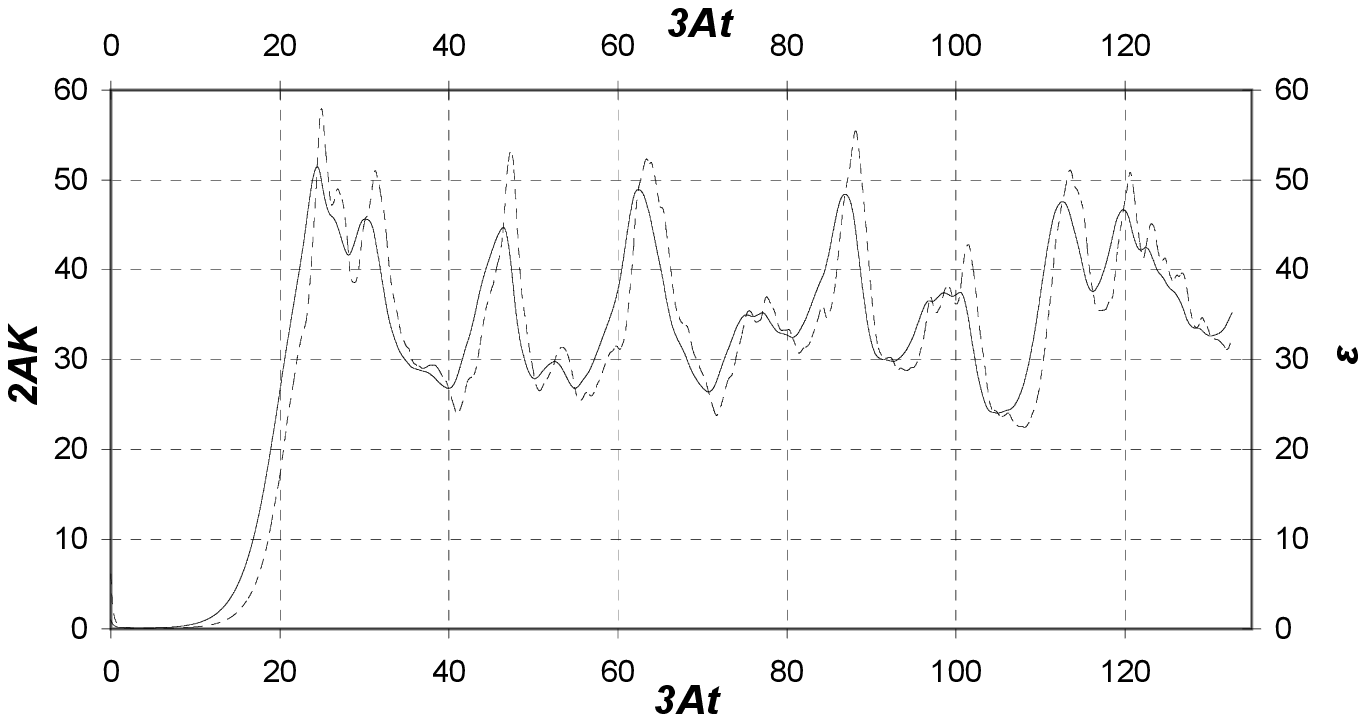}}%
 \\[-3pt]%
  \subfloat[]{\includegraphics[height=0.255\textwidth, angle=0,%
  clip=]{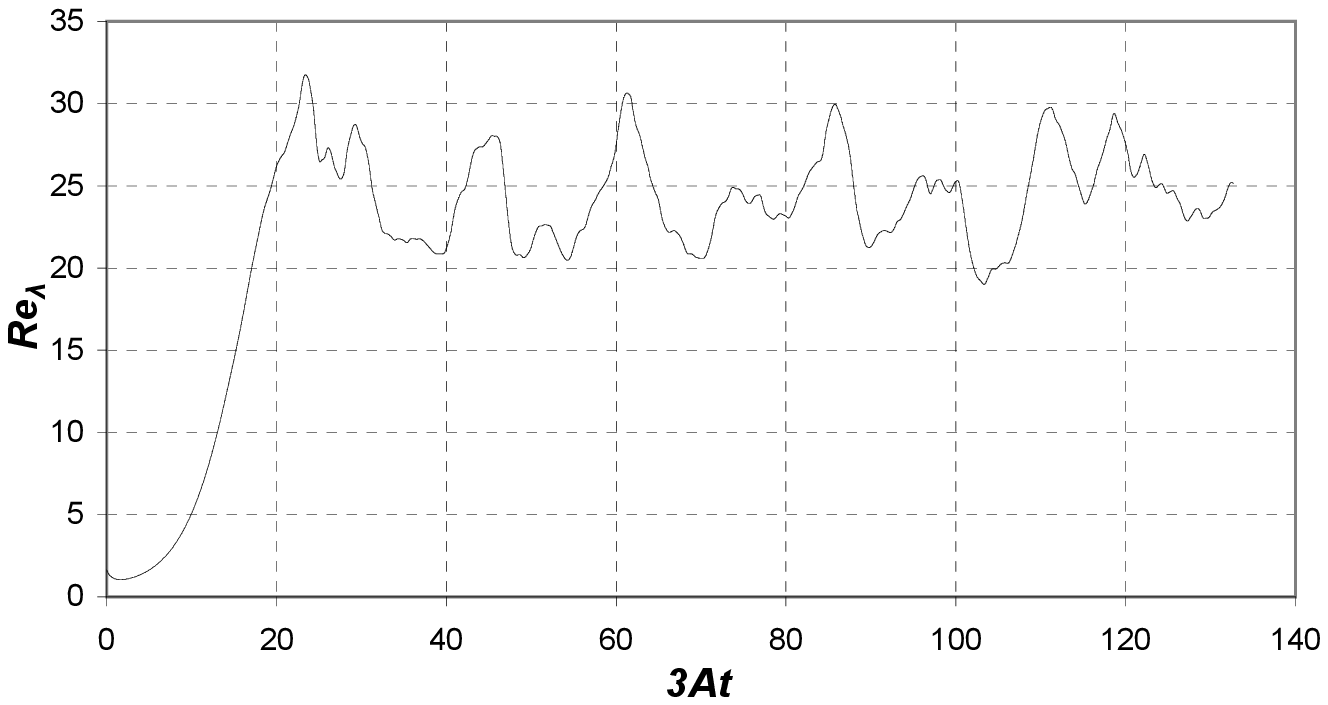}}%
  \end{center}
  \vspace{-0.1in}%
  \caption{A typical evolution of the energy production rate (solid line), dissipation rate (dashed line) is shown in the figure (a) and Taylor microscale Reynolds
  number in figure (b). The parameters chosen are $A=1$, box size $l=2\pi$ and viscosity $\nu=0.1$.}%
  \vspace{-.15in}%
  \label{first-figure}
\end{figure}


Lundgren proposed in \cite{Lund} that we may simplify the
deterministic models to the bare minimum, in some sense, assuming
that the usually velocity dependent force term is merely
proportional to the velocity field for all positions $\mathbf{x}$,
or all wave numbers $\mathbf{k}$, and all times: $\mathbf{f}=A
\mathbf{u}$, where $A$ is plainly a constant. The `linear forcing'
scheme was further studied in \cite{Rosales} and \cite{Akylas1}. Its
simple force term $A \mathbf{u}$ has the same form in both the
spectral and physical space. Thus, unlike other forcing schemes, it
may be used equally easy in cases that need to be solved directly in
the physical space with boundary conditions different than
periodic~\cite{Rosales}. That feature could prove useful.
Additionally, although in the linear forcing the injection of energy
into the flow is not restricted to the larger scales, this scheme
performs decently, and in fact possibly better, in the region
between the inertial range and the integral scale than other forcing
schemes in~\cite{Lund}. From the theoretical point of view what
matters most is that, unlike limited spectral bandwidth forcing
schemes, linear forcing does not introduce an additional length
scale in the problem at the level of the Navier-Stokes equations (a
length scale outside the equations is of course introduced by the
boundary conditions).

The performance of the linear forcing scheme with respect to its
convergence properties was studied in considerable detail in
\cite{Rosales} and useful remarks have been made in
\cite{TurbulenceIII}. The clear conclusion is that linear forcing
results in relatively large fluctuations in the stationary phase.
Indeed, a typical evolution of the energy production rate $2AK$
(where $K$ is the total kinetic energy per unit mass), the
dissipation rate $\varepsilon$ and the Taylor microscale Reynolds
number $\textrm{Re}_\lambda$ is shown in fig.~\ref{first-figure}.
[The details of the DNS can be found in \cite{Akylas1}]. From the
practical point of view this is a disadvantage as it requires longer
simulations in order to obtain good statistics. Also, the stationary
state is reached after a relatively long transient
period~\cite{Rosales}\cite{TurbulenceIII} requiring even more
computational time. On the other hand, linear forcing leads to quite
controllable situations in the stationary state: Given the scales of
the problem i.e.\ the rate $A$, the cubic box size $l$ and the
viscosity $\nu$, the facts of the stationary state are predictable.
The balance between the energy production and dissipation,
$2AK=\varepsilon$, is indeed observed on the (time-)average
validating the very concept of a stationary state; the dissipation
length $L_\varepsilon=(2K)^{\frac{3}{2}}/\varepsilon$ turns out to
be equal to the box size $l$  within few percent error in all
cases~\cite{Rosales}; the Reynolds number
$\textrm{Re}_L=K^2/(\varepsilon \nu)$ may be re-written as
$\frac{1}{4} A L_\varepsilon^2/\nu$ at the stationary state, should
then be roughly equal to $\frac{1}{4}$ of the natural order of
$\textrm{Re}_L$ in this problem, $Al^2/\nu$, in all cases, as it is
indeed observed~\cite{Akylas1}. For example, the Taylor microscale
Reynolds number
$\textrm{Re}_\lambda=(\frac{20}{3}\textrm{Re}_L)^{\frac{1}{2}}$ is
expected to be roughly equal to 25.7 for the run shown in
fig.~\ref{first-figure}. Indeed the average of the
$\textrm{Re}_\lambda$ time-series in fig.~\ref{first-figure} differs
only by few percent from that estimate.

Even if we take stationarity for granted, its characteristics i.e.,
the relatively large fluctuations and the `predictability' of
quantities describing the state of turbulence, certainly call for
understanding. On the other hand the very existence of a stationary
state in this scheme is a fairly intriguing matter. The long-time
effect of the energy production competing with dissipation is not a
priori clear. From the dynamical point of view, it is clear that the
dissipation term $\nu \nabla^2 u$ becomes stronger than the force
term $A u$ at scales smaller than
$(\nu/A)^{\frac{1}{2}}\sim\textrm{Re}_\lambda^{-1} l$, but it is not
clear whether energy which is produced at all other scales up to $l$
will be dissipated by an adequate rate at those smaller scales.

We will approach the problem as follows. The relative simplicity of
linear forcing allows us to study its late-time evolution employing
scaling symmetry arguments to an extent enjoyed possibly only in
freely decaying turbulence; in fact as we shall show there is a
relationship between linearly forced and freely decaying turbulence.
A quite parallel discussion between them can be made. The sections
\ref{Unforced turbulence with decaying viscosity}-\ref{Homogeneous
and isotropic turbulence} will be devoted in presenting these
arguments. The predictability, as we called it above, of the
stationary state, is enlightened through those symmetry arguments,
essentially on the basis that there is no intrinsic large length
scale in the dynamical equations apart from that introduced by the
boundary conditions i.e. the finite size $l$ of the domain. Then
remains the question why the fluctuations observed in the stationary
state, as seen e.g. in the fig.~\ref{first-figure}, are so large. We
shall argue, as analytically as we can, that the fluctuations can be
associated with the deviations from isotropy accumulated by this
forcing at all scales between the scale $(\nu/A)^{\frac{1}{2}}$ and
the domain size $l$ (unlike the limited bandwidth forcing schemes
which feed anisotropy only at the domain size scale where isotropy
is already broken). The method we shall use is to reduce the
dynamical problem to a two-equation model. As a cross check of our
previous conclusions, the stationary state re-emerges as a stable
fixed point of the evolution, a byproduct of which is that
fluctuations tend to be suppressed as long as turbulence is
isotropic. This part of our discussion is presented mostly in
sections \ref{The state of isotropy} and \ref{Self-preserving
turbulence and linear stability of stationarity}. It is interesting
to note that, from various aspects, linearly forced turbulence seems
to be a natural context for the direct application of various ideas
that have been developed in the study of freely decaying turbulence,
in fact one may dare say, an even more natural context.



In terms of equations, a linearly forced incompressible flow with
zero mean flow velocity is described by the Navier-Stokes equation
\begin{equation}\label{1}
\frac{\partial \mathbf{u}}{\partial t}+(\mathbf{u} \cdot \nabla)
\mathbf{u}=-\frac{1}{\rho}\nabla p+\nu \nabla^2 \mathbf{u}+ A
\mathbf{u}\,,
\end{equation}
where the incompressibility condition reads $\nabla \cdot
\mathbf{u}=0$; the velocity field is solenoidal. $A$ is a positive
constant with dimensions of inverse time. The term $A \mathbf{u}$ is
a curious `anti-drag' force on fluid particles. As already mentioned
we impose periodic boundary conditions:
$\mathbf{u}(x,y,z)=\mathbf{u}(x,y+l,z)=\mathbf{u}(x,y,z+l)=\mathbf{u}(x+l,y,z)$.
That is, the flow evolves within a cubic domain with side equal to
$l$ obeying the given conditions on its boundary. (We will often
refer to the cubic domain simply as the `box'.) As we shall
emphasize later on, the term `cubic domain' is slightly misleading
due to the periodic boundary conditions: the flow essentially
evolves in a boundary-less space of finite size. There are no walls
anywhere, this is why we describe the domain as finite instead of
bounded. The problem we are interested in to determine the late-time
state of the turbulent flow governed by these equations and
conditions.

The present work is organized as follows. In sections \ref{Unforced
turbulence with decaying viscosity} and \ref{section exact symmetry}
a reformulation of the problem and an associated scaling symmetry is
presented. In section \ref{Scaling symmetries and asymptotic
behavior} the implications of the scaling symmetry and of the
symmetries of the domain for the late-time behavior of the ensemble
average correlators are discussed. In section \ref{Homogeneous and
isotropic turbulence} we restrict ourselves to isotropic turbulence,
to argue in a more detailed manner for the stationary state as the
final phase of the linearly forced turbulence, as described by the
exact ensemble average correlators and taking into account the
effects of the finiteness of the domain. In section \ref{The state
of isotropy} the expected behavior of the actual observables in DNS
i.e.\ the box-averaged correlators, is discussed in relation to the
properties of the ensemble average correlators established in the
previous sections. In section \ref{Self-preserving turbulence and
linear stability of stationarity} we combine the powerful condition
of isotropy with the (by now established) existence of fluctuations
around stationarity: A complete self-preserving isotropic turbulence
model is obtained and applied to study the fate of fluctuations at
scales in the flow where isotropy holds. We close with a few
additional remarks and discussing certain open issues of the problem
in the final section.

\section{Unforced turbulence with decaying viscosity}
\label{Unforced turbulence with decaying viscosity}

We shall proceed as follows. Mathematically, we may re-write the
problem as an equation for a new field $\mathbf{u}'$ which we
require to satisfy
\begin{equation}\label{2}
\frac{\partial \mathbf{u}'}{\partial t'}+(\mathbf{u}' \cdot \nabla)
\mathbf{u}'=-\frac{1}{\rho}\nabla p'+\nu' \nabla^2 \mathbf{u}'\,,
\end{equation}
with $\mathbf{u}'$ being related to $\mathbf{u}$ by
\begin{equation}\label{3}
\mathbf{u}'=F\, \mathbf{u}\,,
\end{equation}
where $F$ is a function of time $F=F(t')$ to be determined. $\nabla
\cdot \mathbf{u}=0$ implies that $\nabla \cdot \mathbf{u}'=0$.

Substituting (\ref{3}) to (\ref{2}) we get
\begin{equation}\label{}
\frac{1}{F^2}\frac{ dF}{dt'}\mathbf{u}+\frac{1}{F}\frac{\partial
\mathbf{u}}{\partial t'}+(\mathbf{u} \cdot \nabla)
\mathbf{u}=-\frac{1}{F^2}\frac{1}{\rho}\nabla p'+\frac{1}{F}\nu'
\nabla^2 \mathbf{u}\,.
\end{equation}
This equation becomes identical to (\ref{1}) upon setting
\begin{align}
 F dt'=dt\,, \quad
\frac{1}{F^2}\frac{dF}{dt'}=-A\,, \quad  \nu'=F \nu \,.
\end{align}
The transformation of pressure, $p'=F^2 p$, follows from its Laplace
equation constraint and cannot be regarded as an independent
condition.

For constant $A$ we have that
\begin{equation}\label{7}
F=e^{-At}\,,\quad  \textrm{and} \quad At'+1=e^{At}\,,
\end{equation}
where we fix one integration constant by setting $t'=0$
corresponding to $t=0$. It is left understood due to the freedom of
the second integration consttant that we may shift $t$ arbitrarily.

Viscosity $\nu'$ reads
\begin{align}\label{nu'}
\nu'=\frac{1}{At'+1}\nu\,,
\end{align}
in the unphysical time coordinate $t'$. The problem has become
unforced turbulence with decaying viscosity. The way it decays,
$\propto 1/t'$, is crucial in what follows.

Note that everything can be transformed back to the initial
variables at all times, except $t=\infty$, or $t'=\infty$. This is a
singular point of the transformation as $F$ vanishes and the two
forms of the problem are not equivalent. That would be a relevant
subtlety only if we had to deal with the actual limit $t \to
\infty$. We will not need such a limit anywhere in our analysis.

It also worth to note that a transformation $t \to t'$ can generate
only a term which is linear in $\mathbf{u}$, unless the
transformation depends itself on the velocity field. Therefore the
\textsl{possibility} of such a generating transformation is
intimately related to linear forcing.

\section{Time-translation invariance and an exact scaling
symmetry at late times}\label{section exact symmetry}

Note that equation (\ref{1}) does not explicitly depend on time,
forcing being a function of the velocity field alone. Therefore,
depending on boundary conditions, this equation may admit
non-trivial solutions which are static or independent of time in a
certain sense. As we are interested in fully developed turbulence,
the time-dependence in question will apply to statistically defined
quantities.

The Navier-Stokes equations we arrived at by transforming to time
$t'$ in the previous section explicitly reads
\begin{equation}\label{NV'}
\frac{\partial \mathbf{u}'}{\partial t'}+(\mathbf{u}' \cdot \nabla)
\mathbf{u}'=-\frac{1}{\rho}\nabla p'+\frac{\nu}{A t'+1} \nabla^2
\mathbf{u}'\,,
\end{equation}
where $\rho$, $\nu$ and $A$ are constants and $\nabla \cdot
\mathbf{u}'=0$. Apart from the fairly peculiar time-dependence of
viscosity, that is, the energy dissipation rate decreases with time,
the form of this equation is more familiar than that of equation
(\ref{1}).

Now, the time-translation invariance of equation (\ref{1})
translates to an exact scaling symmetry of equation (\ref{NV'}).
Even by inspection one may verify that the transformation
\begin{equation}\label{symmetry}
t' \to e^a t'\,, \qquad \mathbf{u}' \to e^{-a} \mathbf{u}'\,,
\end{equation}
for any constant $a$ is an exact symmetry of the previous equation
(necessarily, $p' \to e^{-2a} p'$) for times $t' \gg A^{-1}$. Then
the integration constants in the relation between $t$ and $t'$ are
irrelevant.

The origin of the late-time scaling symmetry is clear: Shifting $t$
means rescaling $t'$, at least for times $t' \gg A^{-1}$. Shifting
$t$ is a symmetry of equation (\ref{1}), therefore rescaling $t'$
must be a symmetry of equation (\ref{NV'}), as it is indeed the
case. It is important to remember that the symmetry (\ref{symmetry})
respects the periodic boundary conditions on the field
$\mathbf{u}'$, therefore it is an exact symmetry of the problem.

We may now forget equation (\ref{1}) for a little while and focus on
the unforced turbulence described by (\ref{NV'}). Its scaling
symmetry will allow us to draw certain conclusions about the late
time behavior of the system.

\section{Scaling symmetries, asymptotic behavior and isotropy}
\label{Scaling symmetries and asymptotic behavior}

In order to get a first idea why the symmetry can be useful that
way, note that the product $t' \mathbf{u}'$ is invariant under the
scaling (\ref{symmetry}). Consider an arbitrarily chosen moment of
time $t_0'$ and the velocity field $\mathbf{u}_0'$ at that moment,
and another moment $t'=e^a t_0'$ when velocity is $\mathbf{u}'$.
Invariance means: $t' \mathbf{u}'=t_0' \mathbf{u}'_0$. Equivalently
we may write
\begin{equation}\label{u large t}
\mathbf{u}'=\frac{1}{t'}\, t_0'\, \mathbf{u}_0'\,.
\end{equation}
Now in general a symmetry transformation moves us around the space
of solutions. That is, all the previous relation means, is that if
there is a solution with velocity $\mathbf{u}_0'$ at time $t_0'$
then there is another solution with velocity field $\mathbf{u}'$ at
time $t'$. i.e.\ in general $\mathbf{u}'$ and $\mathbf{u}_0'$ need
not necessarily correspond to the same initial conditions.

On the other hand, the symmetry holds for \emph{large} times $t'$
and $t_0'$. Even if it did not, that would be a convenient choice
for the following reason. The initial time $t'=0$ is pushed into the
remote past, and the behavior (\ref{u large t}) might then be an
exact \emph{asymptotic} result for a large class of solutions,
meaning irrespectively of their initial conditions. That implies
that $t' \mathbf{u}'=t_0' \mathbf{u}'_0$ is an actual constant at
each point $\mathbf{r}$ in space depending only on the parameters of
the equation and the boundary conditions.

The constant in question is a vector. To be more specific, recalling
that $\mathbf{u}'$ satisfies $\nabla \cdot \mathbf{u}'=0$, we need a
solenoidal vector field in steady state which does not depend on
initial conditions i.e.\ it is unique. Such a field must respect the
symmetries of the boundary conditions, that is the symmetries of the
cube. There is no such thing: solenoidal vector fields have closed
integral curves which can always be reversed by reflections. We
deduce then that (\ref{u large t}), as long as it is non-trivial,
will always depend to some extend on $t_0'$ i.e.\ on initial
conditions. Thus it is not of much use in this form.

Our reasoning can be used more effectively if it is applied in
statistically defined quantities, that is, correlators of the
velocity field. As mentioned already in the Introduction, from here
and up until section \ref{The state of isotropy} we shall work with
correlators defined as averages over a statistical ensemble. The
statistical ensemble averages are independent of the initial
conditions by their very definition: they are averages over the
space of solutions. Of course in a problem on turbulence they
certainly are the quantities of interest. The statistical ensemble
averages will be denoted by an overbar.

Then (\ref{u large t}) holds trivially for no mean flow:
$\overline{\mathbf{u}'}=0$. Next one considers general correlators
of the velocity field,
$\overline{u'_{i_1}(\mathbf{r}_1,t'_1)u'_{i_2}(\mathbf{r}_2,t'_2)
\cdots}$, and their derivatives. Consider local correlators i.e.\
all times and positions coincide. These are tensor fields $T'_{j_1
j_2\cdots}(\mathbf{r},t')$. Let such a tensor field with $n$
velocity field insertions in the correlation. Then the symmetry
(\ref{symmetry}) tells us, similarly to (\ref{u large t}), that
\begin{equation}\label{T large t}
T'_{j_1 j_2 \cdots}=\frac{1}{t'^n}\, t'^n_0\, T'_{0\, j_1
j_2\cdots}\,.
\end{equation}
Now $t'^n_0\, T'_{0\, j_1 j_2\cdots}(\mathbf{r},t_0')$ must be a
constant at each point $\mathbf{r}$ in space. If not, then this
quantity does depend on $t_0'$ i.e.\ on the  initial conditions.
That means: this quantity is not well defined as a ensemble average
i.e.\ mathematically does not exist and it must be defined in an
approximate manner which does not possess the expected properties,
or only approximately. The reason why this may happen is that the
system has not reached a stage where ensemble averages are
meaningful, a priori some kind of equilibrium is required.

Now, same as with $t_0' \mathbf{u}'_0$, most of these constant
tensor fields must be zero by being inconsistent with the symmetries
of the cube (especially reflections) and the incompressibility
condition. Certainly everything with at least one solenoidal index
must vanish. This leaves us with the scalars, tensors manufactured
out of them and Kronecker delta, and correlators such as
$\overline{\partial_i u_k
\partial_j u_k}$ with no free solenoidal index.

In order to see how these statements are realized by an example,
consider the correlator $t'^2_0\,\overline{u'_{0i}u'_{0k}}$, which
is constant in time. Being constant in time means that it must
respect the symmetries of the cubic domain: it must not change under
reflections of the domain around planes of symmetry and rotations
around axis of symmetry. One should recall that our correlators are
ensemble averages over the whole of phase space, thus symmetries
cannot take us to an other constant late-time solution: there is no
other solution, or we have convergence problems in the very
definition of our averages. It is then easy to see that
$t'^2_0\,\overline{u'_{0i}u'_{0k}}$ must be equal to
$\delta_{ik}\,t'^2_0\,\overline{u'_{0i}u'_{0i}}(\textrm{no
sum})=\frac{1}{3}\delta_{ik}\,t'^2_0\,\overline{u'_{0j}u'_{0j}}$,
i.e.\ essentially a scalar. Moreover, by the incompressibility
condition $\nabla \cdot \mathbf{u}'=0$ we see that the scalar itself
must be constant in space.

One should note that the situation resembles very much that of
isotropic i.e.\ also homogeneous turbulence. There is anisotropy
allowed by the problem but it is much less than what would call
anisotropy in general. Thus we will proceed by assuming isotropy and
analyze what that implies; then, as isotropy cannot hold at scales
comparable to the cubic box size $l$, the effects of the boundary
eventually play a key role. This is done in the next section. We
close this section by defining a few important scalars for the
description of turbulence, their symmetry and transformation
properties and their expected late time behavior according to our
arguments.

The r.m.s. value $q$ of the velocity and the dissipation rate
$\varepsilon$ are defined by $q^2=\overline{\mathbf{u}\cdot
\mathbf{u}}$ and $\varepsilon=\nu\, \overline{\partial_j u_i
\partial_j u_i}$. Also by $K=\frac{1}{2}q^2$ we shall denote the total kinetic energy per unit mass.
Similar expressions hold for the primed quantities.

Under the symmetry (\ref{symmetry}) the quantities $K'$ and
$\varepsilon'$ transforms as
\begin{equation}\label{}
K' \to e^{-2a} K'\,, \qquad \varepsilon' \to e^{-3a} \varepsilon'\,,
\end{equation}
where one should bear in mind that $\varepsilon'$ involves $\nu'$
defined in (\ref{nu'}). Following again the reasoning given in the
previous paragraphs we conclude that for large times $t'$ the
kinetic energy and dissipation rate should obey
\begin{equation}\label{K' of t'}
K'=\frac{\textrm{constant}}{t'^2}\,,\qquad
\varepsilon'=\frac{\textrm{constant}}{t'^3}\,.
\end{equation}

In order to see what this result means back in the variables of the
system (\ref{1}), we use (\ref{3}) and (\ref{nu'}) to obtain the
transformations of $K$ and $\varepsilon$:
\begin{equation}\label{K' to K}
K'=(At'+1)^{-2}K\,, \qquad \varepsilon'=(At'+1)^{-3}\varepsilon\,.
\end{equation}
The result is then that the kinetic energy and dissipation rate in
the linearly forced turbulence should at late times become
\begin{equation}\label{K of t}
K=\textrm{constant}\,, \qquad \varepsilon=\textrm{constant}\,.
\end{equation}
Presumably, the dissipation length scale $L_\varepsilon$ and the
Reynolds number $\textrm{Re}_L$ defined by
\begin{equation}\label{Le}
L_\varepsilon=\frac{q^3}{\varepsilon}\,, \qquad
\textrm{Re}_L=\frac{K^2}{\varepsilon\nu}\,,
\end{equation}
and transforming by
\begin{equation}\label{Re}
{\textrm{Re}'}_{\!\!L}=\textrm{Re}_L\,, \qquad
{L'}_{\!\!\varepsilon}=L_\varepsilon\,,
\end{equation}
should also reach constant values. That is, turbulence should get to
what we have already called as the stationary state or phase.

\begin{center} *** \end{center}


The arguments given above can be rephrased in the actual time $t$
and the variables of equation (\ref{1}) as follows. We have already
mentioned that shifting time $t$ is a symmetry of equation
(\ref{1}). That is, if $\mathbf{u}(t)$ is a solution of this
equation then so is $\mathbf{u}(t+\Delta t)$ for an arbitrary
interval $\Delta t$. These two solutions do not coincide because
they correspond to the different initial conditions. On the other
hand, we may say that for a certain class of initial conditions,
that difference should become irrelevant at late times i.e.\ the two
solutions, or at least certain quantities calculated out of them,
will coincide. But this is the same as stating the obvious fact that
static or stationary solutions of (\ref{1}) exist, without
explaining whether such stationary states are indeed the ending
point of solutions for an reasonably large class of initial
conditions. In this light our arguments as given so far seem rather
trivial.

Our arguments are essentially about symmetries. The most convenient
context to discuss them, and possibly the only context, is that of
isotropic turbulence. We shall argue that the evolution of the
system to the stationary phase, can be thought of as the gradual
breaking of a larger approximate symmetry to the smaller exact
symmetry (\ref{symmetry}), which is solely consistent with the
stationary state. That will be realized in certain convenient cases
where one may convince oneself that one `watches' the system
evolving as claimed. By the very form of (\ref{NV'}) one may guess
that standard knowledge from the freely decaying turbulent flows
could prove useful to us.

We start by reviewing certain useful facts about the freely decaying
isotropic turbulence.

\section{Homogeneous and isotropic turbulence}
\label{Homogeneous and isotropic turbulence}

\subsection{Important quantities and formulas}
\label{Important quantities and formulas}

Consider homogeneous and isotropic turbulence. The one-direction
r.m.s. value of the velocity, $q_1$, does not depend on the
direction, i.e.\, $q^2=3 q_1^2$. The two-point correlation function
of the velocity is reduced to a scalar $f(r)$ which depends only the
distance $r$ between the two points:
$\overline{u_l(0)u_l(\mathbf{r})}=q_1^2\, f(r)$. The entire
information of the two-point correlation is contained in components
$u_l$ longitudinal in the direction of separation. Also the
two-point triple correlation of the velocity can only have
longitudinal components and it is expressed in terms of a scalar
$h(r)$ by $\overline{u_l(0)u_l(0)u_l(\mathbf{r})}=q_1^3\, h(r)$. A
priori all quantities depend on time, for that reason
time-dependence is left understood.

Equation (\ref{1}) with $A=0$ is the unforced Navier-Stokes equation
describing turbulence in the freely decaying state. The
`Karman-Howarth equation' \cite{LandauFluidMechanics}\cite{Frisch}
derived from it under the conditions of homogeneity and isotropy
reads
\begin{equation}\label{KH}
\frac{\partial}{\partial t}(q_1^2
f)=\frac{1}{r^4}\frac{\partial}{\partial
r}\Big\{r^4\Big(q_1^3\,h+2\nu\, q_1^2\,\frac{\partial f}{\partial
r}\Big)\Big\}\,.
\end{equation}

In freely decaying turbulence the rate at which energy $K$ is
decreasing equals the dissipation rate $\varepsilon$, expressing the
balance of total energy in that problem.
\begin{equation}\label{energy balance free}
\dot K=-\varepsilon\,.
\end{equation}
Presumably, this also holds if the viscosity $\nu$ depends
explicitly on time. This fact will be useful below.

The integral scale, $L \equiv \int_0^\infty \! f dr$, is of the
order of magnitude of the dissipation length $L_\varepsilon$. The
Taylor micro-scale $\lambda_g$ is defined by a differential relation
involving $f$:
\begin{equation}\label{taylor lambda general}
\lambda_g^2\left.\frac{\partial^2f}{\partial
r^2}\right|_{r=0}\!\!\!=-1\,, \quad \textrm{i.e.\ } \quad
\lambda_g=\sqrt{\frac{10 \nu K}{\varepsilon}}\,.
\end{equation}

For completeness, and as we shall briefly need it later on, we write
down the energy balance equation for the spectral densities of $K$
and $\varepsilon$. It is a Fourier transform of the Karman-Howarth
equation (\ref{KH}), see e.g.\ \cite{Frisch}.
\begin{equation}\label{energy balance spectral}
\partial_t E(k)=-\partial_k  T(k)-2\nu k^2 E(k)\,.
\end{equation}
The `spectrum' $E(k)$ suitably integrates to give the kinetic energy
and dissipation rate, $K=\int_0^\infty E(k) dk$ and
$\varepsilon=2\nu \int_0^\infty k^2 E(k) dk$. $T(k)$ is the spectral
energy flux and vanishes for vanishing and infinite wave numbers.
Clearly (\ref{energy balance free}) follows by integrating
(\ref{energy balance spectral}) over all $k$, though the derivation
of energy balance equations will be discussed in more detail section
\ref{The state of isotropy}.

\subsection{Scaling symmetries and power laws}
\label{Scaling symmetries and power laws}

The scaling arguments given in this section are borrowed from
\cite{Oberlack1}. The method is an application of the reasoning
presented in section \ref{Scaling symmetries and asymptotic
behavior}.

Consider the Karman-Howarth equation (\ref{KH}). Now perform the
two-parameter scaling transformation
\begin{align}\label{a b transformation}
& t \to e^a t\,, \nonumber\\
&r \to e^b r \,, \nonumber\\
& q \to e^{b-a} q \,, \\
& f \to f\,, \nonumber\\
& h \to h\,, \nonumber
\end{align}
for arbitrary constants $a$ and $b$. Changing $a$ for fixed $t$
amounts to time evolution from the initial moment $t$. Similarly
changing $b$ for fixed $r$ amounts to looking at larger distances.
Under (\ref{a b transformation}) equation (\ref{KH}) becomes
\begin{equation}\label{KH-transformed}
\frac{\partial}{\partial t}(q_1^2
f)=\frac{1}{r^4}\frac{\partial}{\partial
r}\Big\{r^4\Big(q_1^3\,h+e^{a-2b}\,2\nu\, q_1^2\,\frac{\partial
f}{\partial
r}\Big)\Big\}\,.
\end{equation}
Consider high Reynolds numbers. Then the viscosity term can be
dropped. We see then that the transformation (\ref{a b
transformation}) is an \emph{approximate} symmetry of (\ref{KH}) for
high Reynolds numbers; it can be regarded as a symmetry of the
system for infinite Reynolds numbers.

Consider then quantities of interest such as the kinetic energy $K$,
or the integral scale $L$ (equivalently, the dissipation length
$L_\varepsilon$). They transforms same as $q^2$ and $r$
respectively.

The one-parameter subgroup of the transformation (\ref{a b
transformation}) such that
\begin{equation}
\gamma=\frac{b}{a}\,,
\end{equation}
is an arbitrary but fixed number, given explicitly by
\begin{align}\label{gamma transformation}
& t \to e^a t\,, \nonumber\\
&r \to e^{\gamma a} r \,, \nonumber\\
& q \to e^{\gamma a-a} q \,, \\
& f \to f\,, \nonumber\\
& h \to h\,, \nonumber
\end{align}
for arbitrary $a$, leaves the quantities
\begin{equation}\label{invariant}
t^{-\gamma}L\,, \qquad  t^{2-2\gamma} K\
\end{equation}
invariant.

Note that this way, we think of the two-parameter group (\ref{a b
transformation}) as one-parameter ($\gamma$) family of one-parameter
subgroups (\ref{gamma transformation}). Presumably, equation
(\ref{KH-transformed}) becomes identical to (\ref{KH}) iff $a-2b=0$
that is $a-2\gamma a=0$. This means that the subgroup
$\gamma=\frac{1}{2}$ is an exact symmetry of the freely decaying
turbulence. In other words, the larger symmetry (\ref{a b
transformation}) for infinite Reynolds number breaks down to its
subgroup $\gamma=\frac{1}{2}$ for finite Reynolds numbers.

Each symmetry (\ref{gamma transformation}) is essentially
time-evolution. Following the arguments of section \ref{Scaling
symmetries and asymptotic behavior} we conclude that at adequately
late times
\begin{equation}\label{L K of gamma}
L= \textrm{constant}\,\, t^{\gamma}\,, \qquad
K=\textrm{constant}\,\, t^{2\gamma-2}\,.
\end{equation}
Thus we have obtained certain power laws for the late behavior of
the length scale and kinetic energy in freely decaying turbulence.
The law for the dissipation rate $\varepsilon$ follows immediately
by (\ref{energy balance free}),
\begin{equation}\label{epsilon of gamma}
\varepsilon=\textrm{constant}\, t^{2\gamma-3}\,.
\end{equation}
Then, by (\ref{Le}) the law for dissipation length $L_\varepsilon$
turns out consistent with that of $L$, as course it should. The law
for $\textrm{Re}_{L}$ also follows from (\ref{Le}):
\begin{equation}\label{Re of gamma}
\textrm{Re}_L=\textrm{constant}\, t^{2\gamma-1}\,.
\end{equation}


Summarizing, each value of $\gamma$ defines a subgroup of the full
symmetry group (\ref{a b transformation}) for high Reynolds. Given a
$\gamma$ the time-dependence of various quantities takes the form of
specific power laws. A priori not fixed without additional
conditions, the exponent $\gamma$ may be given an additional
physical interpretation. Assume that for low wave-numbers $k$ the
spectrum $E(k)$ is of the form
\begin{equation}
E(k)=C k^\sigma+o(k^\sigma)\,,
\end{equation}
for some \emph{constants} $C$ and $\sigma$. Given the dimension of
$E(k)$ and $k$ and the constancy of $C$, this relation is invariant
under (\ref{gamma transformation}) iff
\begin{equation}\label{gamma-sigma}
\gamma=\frac{2}{\sigma+3}\,.
\end{equation}
That is, the subgroup (\ref{gamma transformation}) is fixed by the
small wave-number behavior of the spectrum of the specific class of
flows. It may be argued, see e.g.\ Ref.~\cite{Lesieur}, that $C$ is
actually constant as long as $1 \le \sigma < 4$; also the case
$\sigma=4$ holds marginally. That is, in those cases $C$ is fixed by
the initial conditions.


Decay exponents are usually expressed in terms of $n \equiv
2-2\gamma$, which is the kinetic energy decay exponent, $K \sim
t^{-n}$. About the value of $n$ there are well known suggestions.
They depend upon the identification of $C$ with quantities which are
conserved under certain conditions. Kolmogorov~\cite{K41-decay} and
Batchelor~\cite{Batchelor48}, based on the conservation of the
Loitsyanky integral~\cite{Loitsianski}, derived $\gamma=\frac{2}{7}$
i.e.\ $n=\frac{10}{7}$. Saffman~\cite{Saffman1} set forth the
hypothesis that the vorticity and not the velocity correlator is an
analytic function in spectral space, by which rediscovered the
$\sigma=2$ spectrum and Birkhoff's
integral~\cite{Birkhoff-Turbulence} and derived $\gamma=\frac{2}{5}$
i.e.\ $n=\frac{5}{6}$.
Experimentally~\cite{Comte-Bellot-Corssin}\cite{Warhaft-Lumley} both
values of the decay exponent $n$ are acceptable. The value $n=1$ has
also been suggested by other theoretical considerations, for high
but finite Reynolds numbers~\cite{Speziale} and as the limiting
value of the decay exponent for infinitely high Reynolds
numbers~\cite{george-92:1492,Barrenblatt-book,lin}; this solution
first appeared in \cite{dryden}.

The $n=1$ decay solution for finite Reynolds numbers of
Ref.~\cite{Speziale} can be obtained by recalling an observation
given above, that for finite Reynolds numbers the symmetry (\ref{a b
transformation}), essentially associated with infinite Reynolds
numbers, breaks down to its $\gamma=\frac{1}{2}$ subgroup at finite
Reynolds numbers. That means $n=1$. Presumably, by (\ref{Re of
gamma}), the Reynolds number is constant for this solution.

The $n=1$ decay law may also be obtained in another way, which gives
the chance to make an additional comment on the analysis presented
in this section. The Taylor microscale $\lambda_g$ transforms as a
length, same as $r$, according to the equation on the left in
(\ref{taylor lambda general}). That means that $\lambda_g =
\textrm{constant}\, t^{\gamma}$. On the other hand the equation on
the right in (\ref{taylor lambda general}) and the laws (\ref{L K of
gamma}) and (\ref{epsilon of gamma}) imply
$\lambda_g=\textrm{constant}\, t^{\frac{1}{2}}$. The reason why
there is no discrepancy is because regarding $L$ as finite and
Reynolds number virtually infinite, for the symmetry (\ref{a b
transformation}) to hold, means that $\lambda_g$ is virtually zero.
Put differently, if we want to think of the previous analysis as
applying also to high but finite Reynolds numbers, then we must
restrict ourselves to scales much larger than Taylor microscale. It
is then no accident that the power laws (\ref{L K of gamma}) can
also be produced by models deriving from self-similarity of
turbulence with respect to the integral scale $L$, as we shall
discuss in section \ref{Self-preserving turbulence and linear
stability of stationarity}. On the other hand if we want finite
Reynolds \emph{and} to take into account scales of $O(\lambda_g)$ or
less, then it must be $\gamma=\frac{1}{2}$ i.e.\ $n=1$.

Being such a direct implication of the arguments in this section,
one may wonder why the $n=1$ decay solution is not observed
experimentally even for the highest Reynolds numbers (equivalently,
as $\gamma=\frac{1}{2}$ means $\sigma=1$, a small wave-number
spectrum $E(k) \propto k$ has not been verified). The arguments
possibly fail on the very part where one expects independence from
the initial conditions. That expectation might be in a better shape
the higher, still finite, the Reynolds number is. This is why in the
best case the $n=1$ solution can possibly be regarded as describing
well decaying turbulence for very high Reynolds numbers.

In the next two subsections we come to the problem of interest. The
discussion parallels in some sense our previous remarks: Going from
the infinitely high to any lower Reynolds number the larger symmetry
(\ref{a b transformation}) breaks in this case down to its exact
subgroup $\gamma=0$ associated with the linearly forced turbulence,
the exact symmetry (\ref{symmetry}) we started our discussion with.
But, unlike the freely decaying case, in our problem a large length
scale and a Reynolds number scale are \emph{necessarily} present,
eventually forcing the system towards the $\gamma=0$ evolution. That
amounts to reaching the stationary state.


\subsection{Linearly forced isotropic turbulence}
\label{Linearly forced isotropic turbulence}


Consider linearly forced turbulence in the description given by
equation (\ref{NV'}), which let us state again:
\begin{equation*}
\frac{\partial \mathbf{u}'}{\partial t'}+(\mathbf{u}' \cdot \nabla)
\mathbf{u}'=-\frac{1}{\rho}\nabla p'+\frac{\nu}{A t'+1} \nabla^2
\mathbf{u}'\,.
\end{equation*}
The analogue of the transformed Karman-Howarth equation
(\ref{KH-transformed}) for late times $t' \gg A^{-1}$ reads now
\begin{equation*}\label{KH-transfo-Linear}
\frac{\partial}{\partial t'}(q'^2_1
f')=\frac{1}{r^4}\frac{\partial}{\partial r}\Big\{r^4\Big(q'^3_1 h'+
e^{-2\gamma a}\frac{\nu}{At'} 2q'^2_1\,\frac{\partial f'}{\partial r}\Big)\Big\}\,.
\end{equation*}
We find of course again that the subgroup $\gamma=0$ of the group
(\ref{a b transformation}) i.e.\ the group (\ref{symmetry}), is an
exact late-time symmetry of the linearly forced turbulence. For very
high Reynolds numbers the larger group (\ref{a b transformation}) is
a good approximate symmetry of the system. The evolution laws for
the dimensionful quantities are necessarily similar to those of the
freely decaying case,
\begin{align}\label{L' K' e' of gamma}
& L'= \textrm{constant}\,\, t'^{\gamma}\,,
\\
& K'=\textrm{constant}\,\, t'^{2\gamma-2}\,, \nonumber
\\
& \varepsilon'=\textrm{constant}\, t'^{2\gamma-3}\,. \nonumber
\end{align}
while the dimensionless Reynolds number has a different power law
due to the time-varying $\nu'$:
\begin{equation}\label{Re' of gamma}
\textrm{Re}'_L=\textrm{constant}\, t'^{2\gamma}\,.
\end{equation}

Consider a flow that starts off with velocities of order $u_0$ and a
box of size $l$ such that $u_0 \gg A l$. Equivalently the turn-over
time is much smaller than forcing time scale $A^{-1}$, that is,
$l/u_0 \ll A^{-1}$.

Given $A$, $l$ and $\nu$ there is a naturally defined Reynolds
number in the problem:
\begin{equation}\label{reynolds at stationarity}
\textrm{Re}_{A}=\frac{Al^2}{\nu}\,.
\end{equation}
That is, the condition $l/u_0 \ll A^{-1}$ can be rephrased as that
the flow starts off with a very high Reynolds number, $\textrm{Re}_L
\gg \textrm{Re}_A$.

Consider then times $t'$ such that $l/u_0 \ll t' \ll A^{-1}$.
Looking at the previous equation we understand that for those times
the turbulent flow is merely freely decaying with constant viscosity
$\nu$. If all previous inequalities hold strongly enough, then there
will be time for the flow to evolve adequately towards its developed
stage. That means that the quantities describing turbulence will
evolve according to the power laws (\ref{L' K' e' of gamma}) and
(\ref{Re' of gamma}).


When $t'$ becomes of order of $A^{-1}$ `linear forcing' kicks in.
Now one should recall that Reynolds number is always decreasing,
therefore some time before or after that moment it will drop enough
so that the viscosity term cannot be neglected. That means that from
that moment on the group (\ref{a b transformation}) is not much of a
symmetry anymore: the only symmetry remaining is its subgroup
(\ref{symmetry}) corresponding to $\gamma=0$, which is exact and
therefore holds at all times. Viscosity now decreases with time
therefore energy will be dissipated with an ever decreasing rate. We
may then picture, very roughly, the flow evolving by going through
stages of smaller exponents $\gamma$, following simple of less
simple laws parameterized by it, eventually reaching the specific
value for which a subgroup (\ref{gamma transformation}) is an exact
symmetry of the system: $\gamma=0$.

Let us recapitulate. There is a symmetry existing in the system for
high Reynolds numbers. A priori allows for arbitrary values of the
parameter $\gamma$. This symmetry breaks down to its subgroup
$\gamma=0$ when Reynolds drops enough. This is an exact symmetry of
the system, therefore holds always. The final state must be the one
respecting that exact symmetry. The power laws (\ref{L' K' e' of
gamma}) and (\ref{Re' of gamma}) imply that $L'$ and
$\textrm{Re}_L'$ are constant, $\varepsilon'=\textrm{constant}\,
t'^{-3}$ and $K'=\textrm{constant}\, t'^{-2}$. The transformations
(\ref{K' to K}) and (\ref{Re}) back to the original variables of
equation (\ref{1}) show that everything, that is $K$, $\varepsilon$,
$L_\varepsilon$ and $\textrm{Re}_L$, is constant. We have reached
the stationary state. $K'$ and $\varepsilon'$, which are
time-dependent in the description (\ref{NV'}) of the problem, are
related by the equation (\ref{energy balance free}) written down for
the primed quantities:
\begin{equation}\label{}
\frac{dK'}{dt'}=-\varepsilon'\,.
\end{equation}
The (\ref{L' K' e' of gamma}) and (\ref{Re' of gamma}) power laws
for $\gamma=0$ and the transformations (\ref{K' to K}) translate
that relation to
\begin{equation}\label{}
2AK=\varepsilon\,.
\end{equation}
That is, at the stationary state energy production balances exactly
dissipation.

If initially Reynolds number is not very high, there will be a
strong mixing of the phases of freely decaying and `linear forced'
turbulence. The transition between them is then too complicated to
describe. Though one cannot argue against the possibility the system
having an entirely different behavior, it seems reasonable that the
same mechanisms which lead the system to the power laws consistent
with $\gamma=0$ will do the same thing, in far more complicated way.

\subsection{Effects of the finite domain}
\label{Effects of the finite domain}

A very amusing thing we observe is that, by (\ref{gamma-sigma}), the
value $\gamma=0$ corresponds to $\sigma=\infty$. This indicates that
the power law behavior of the spectrum $E(k)$ for small wave-numbers
degenerates, and should be replaced by some other, much faster
decreasing law, perhaps some kind of exponential.

There is a good reason why we expect that. The system (\ref{1}), or
(\ref{NV'}), is solved in a domain of some finite size $l$. An
infinite size $l$ is meaningless: In the absence of another large
length scale, this means that the total energy production rate in
the domain depends on it and diverges\footnote{The density $\rho$ is
regarded as constant, which for infinite $l$ means also infinite
mass $\rho l^3$. But that it is not a priori harmful, because what
matters is the largest scales where turbulent motions correlate and
the total energy produced in the system.}, $2A K \rho l^3\to
\infty$. Also large $l$ essentially means a large $A l$ compared
with any specific initial condition $u_0$: an infinitely large $l$
is equivalent to initial conditions $u_0$ infinitely close to zero.
Now finite size means that there are no wave-numbers $k$ between
$O(l^{-1})$ and zero, therefore any continuous approximation of the
spectrum $E(k)$ must fall very rapidly for $kl$ smaller than $O(1)$.

There is a major implication following the presence of a finite size
domain. \emph{Its fixed size $l$ breaks the symmetry (\ref{a b
transformation}), as the presence of a fixed length says that it
must be $b=0$.} That is, the domain size breaks the larger symmetry
(\ref{a b transformation}) down to its subgroup $\gamma=0$, the
exact symmetry.

We may now think of the evolution of the flow from another point of
view, that of the integral scale. As long as the integral scale $L'$
is small compared to $l$ the group (\ref{a b transformation}) is a
fairly good approximate symmetry. Then $L'$ increases with time as
$\sim t'^{\gamma}$. As $L'$ grows larger, (\ref{a b transformation})
is a less and less good approximate symmetry. As before, we may then
roughly picture the flow as going through stages of smaller $\gamma$
reaching the stage with $\gamma=0$ which is consistent with the
exact symmetry. This means that $L'$ will become constant. The
natural order for that constant $L'$ (as well as $L$, recalling that
$L=L'$ by the transformation (\ref{Re})) is the domain size $l$. As
mentioned in the Introduction, DNS have shown that specifically
$L_\varepsilon=l$ within a few percent error~\cite{Rosales}.

We may note that for high Reynolds numbers the viscosity term can be
anyway neglected. Therefore the (\ref{L' K' e' of gamma}) and
(\ref{Re' of gamma}) power laws for $\gamma=0$ and the arguments
given above, make sense for very high Reynolds numbers for freely
decaying turbulent flows in a finite domain. The difference in the
`linearly forced' case (\ref{NV'}) is that those power laws are
associated with an exact symmetry of the equations and of course
hold also for lower Reynolds numbers. In contrast, as we have seen
in section \ref{Scaling symmetries and power laws}, the exact
symmetry in the freely decaying turbulence requires
$\gamma=\frac{1}{2}$. Linearly forced turbulence  behaves in a simpler way 
than the freely decaying, in this sense.

One should note that by the time linear forcing kicks in, the
time-dependence of viscosity $\nu'$ changes the form of the Reynolds
number power law. During times that linear forcing is already at
work but the Reynolds number is still very high compared to
$\textrm{Re}_A$ and decreasing, the Reynolds number is given by the
power law (\ref{Re' of gamma}) and decreases due to a decreasing
$\gamma$ as we argued in the previous section. That is, when
$\gamma$ eventually vanishes the peculiar `decaying' turbulent flow
(\ref{NV'}) reaches a peculiar kind of stationarity: its Reynolds
number becomes constant i.e.\ \emph{turbulence as such is not
decaying at all}. This reflects of course the association of this
state to an scaling exact symmetry of the full viscous equations.
Recalling transformation (\ref{Re}), $\textrm{Re}_L'=\textrm{Re}_L$,
this is also the Reynolds number of the linearly forced turbulence
described in the variables of (\ref{1}).



That final value of the Reynolds number can be easily estimated by
the existence of a natural scale $\textrm{Re}_A$, defined in
(\ref{reynolds at stationarity}). At the stationary state energy
production balances dissipation, $2AK=\varepsilon$, which implies
that $\textrm{Re}_L=AL_\varepsilon^2/(4\nu)$ i.e. $\textrm{Re}_A$
indeed sets the scale for the Reynolds number at the stationary
state. In fact, as mentioned in the Introduction, DNS show that
$\textrm{Re}_L =\frac{1}{4} \textrm{Re}_A$ within a few percent
error~\cite{Akylas1}.


We conclude that the finiteness of the domain emerges as a crucial
factor in understanding the flow evolving to a stationary state. Now
the domain, we often referred to as the `box', should be
understood more carefully in terms of periodicity. 
This is what we discuss next.



\section{The state of isotropy}
\label{The state of isotropy}

In the previous section we restricted ourselves to flows obeying the
conditions of homogeneity and isotropy. Let us review what that
involves. Contract (\ref{1}) with $\mathbf{u}$ and average. After a
little re-arranging we have
\begin{align}\label{energy balance locally}
& \frac{\partial K}{\partial t}=-\varepsilon+ 2A K++\nu \nabla^2
K-\nabla \cdot \mathbf{J}\,, \\
& \textrm{J}_i \equiv
\overline{\Big(\frac{u^2}{2}+\frac{p}{\rho}\Big)u_i}\,. \nonumber
\end{align}
Homogeneity alone makes all locally defined correlators, such as $K$
or $J_i$, independent of the position in space. That is, the last
two terms in the last equation vanish. Of course everywhere isotropy
means homogeneity, so if the flow is assumed isotropic those two
terms go again away. We are then left with
\begin{equation}\label{energy balance local homogeneity}
\dot K=-\varepsilon+2AK\,.
\end{equation}
In a stationary state we get $\varepsilon=2AK$, whose form we
already anticipated on dimensional grounds deriving the final value
(\ref{reynolds at stationarity}) of the Reynolds number in terms of
the box size $l$.

We should now recall that the box is a cubic domain together with
periodic boundary conditions we impose on all fields. The periodic
boundary conditions can be given some enlightening interpretations.
One way to think of them is that we solve the Navier-Stokes in an
infinite medium imposing periodicity $l$ on the field
$\mathbf{u}(x,y,z)$ in all three directions $x$, $y$ and $z$. That
in turn means that homogeneity is not a priori broken: a box such
that $\mathbf{u}(x,y,z)$ satisfies periodic boundary conditions can
be drawn anywhere in the infinite medium.  By the periodicity we
apparently restrict ourselves to special kinds of flow such that
there is an upper bound to the size of eddies or any spatially
periodic structure in it.

Another way to think of the periodic boundary conditions is to
interpret them as mere single-valued-ness of the fields while
identifying the points of the boundary of the cube where the fields
are supposed to be equal. This may be pictured if we go one
dimension down. If we take a square and identify the opposite sides
we get a topological torus. A torus is a perfectly homogeneous space
without boundary which is non-trivial globally and it is not
isotropic. Imposing periodic boundary conditions on the cube means
we essentially solve Navier-Stokes equations on the
three-dimensional analogue of such a space, the three-torus.

An implication of periodicity and its peculiar nature is that an
analogue of the equation (\ref{energy balance local homogeneity})
holds without assuming point-wise homogeneity of the flow. Let us
denote by $\langle X\rangle$ the spatial average over the volume of
the box of a quantity\footnote{That is, for example, $\langle K
\rangle$ means simply $\langle\frac{1}{2}\mathbf{u} \cdot
\mathbf{u}\rangle$, not $\langle\frac{1}{2}\overline{\mathbf{u}
\cdot \mathbf{u}}\rangle$. This is a little confusing but allows for
a more compact notation.} with an ensemble average $X$.
Averaging that way we get an equation similar to (\ref{energy
balance locally}) for box-averages
\begin{align}\label{energy balance locally averaged}
\frac{d\langle K \rangle}{d t}=&-\langle\varepsilon\rangle+ 2A
\langle
K\rangle\nonumber+\\&+\frac{1}{V_{\textrm{box}}}\int_{\textrm{bdy}}\{\nu
\nabla K- \mathbf{J}\} \cdot d\mathbf{a} \,.
\end{align}

The last term is a surface integral over the boundary surface of the
box. This integral is a sum of
\begin{equation}\label{pair}
\int_{x=l} (\nu \partial_x K- J_x) dy dz-\int_{x=0} (\nu \partial_x
K- J_x) dy dz
\end{equation}
plus two more analogous pairs of terms for the other two directions.
As all fields are manufactured out of correlations of the field
$\mathbf{u}$ which satisfies periodicity
$\mathbf{u}(x,y,z)=\mathbf{u}(x+l,y,z)$, and the same for the other
two directions, any pair of terms such as (\ref{pair}) vanishes
exactly and identically. In other words even if point-wise
homogeneity is not assumed, periodicity implies that
\begin{align}\label{energy balance locally averaged periodicity}
\frac{d\langle K \rangle}{dt}=-\langle\varepsilon\rangle+ 2A \langle
K\rangle
\end{align}
must hold exactly.

Alternatively, the vanishing of the boundary term in (\ref{energy
balance locally averaged}) follows automatically under the
interpretation that we solve our problem on a three-torus: there
simply is \emph{no boundary}, $x=0$ and $x=l$ describe the
\emph{same} surface somewhere on the three-torus. Whatever the
interpretation of the boundary conditions, one ends up with
(\ref{energy balance locally averaged periodicity}) without assuming
homogeneity of the flow.

This is a good thing to know. The box-averaged correlators $\langle
X\rangle$ are the actual observables in the DNS. Now, what is their
relation to the ensemble averages $X$? Assuming that $X$ are
meaningful and under an ergodic hypothesis, $X$ are represented by
averaging $\langle X\rangle$ over suitable and adequately large
intervals of time. That first of all means that although the motion
of $\langle X\rangle$ is not the same as that of $X$ it should
nonetheless be bounded and appear as fluctuating around hypothetical
stationary values. This is what it is observed,
fig.~\ref{first-figure}. Those values should of course be the values
of the correlators $X$.
\begin{figure}[t]\vspace{.0in}
\begin{center}
  \includegraphics[height=0.28\textwidth,  angle=0]{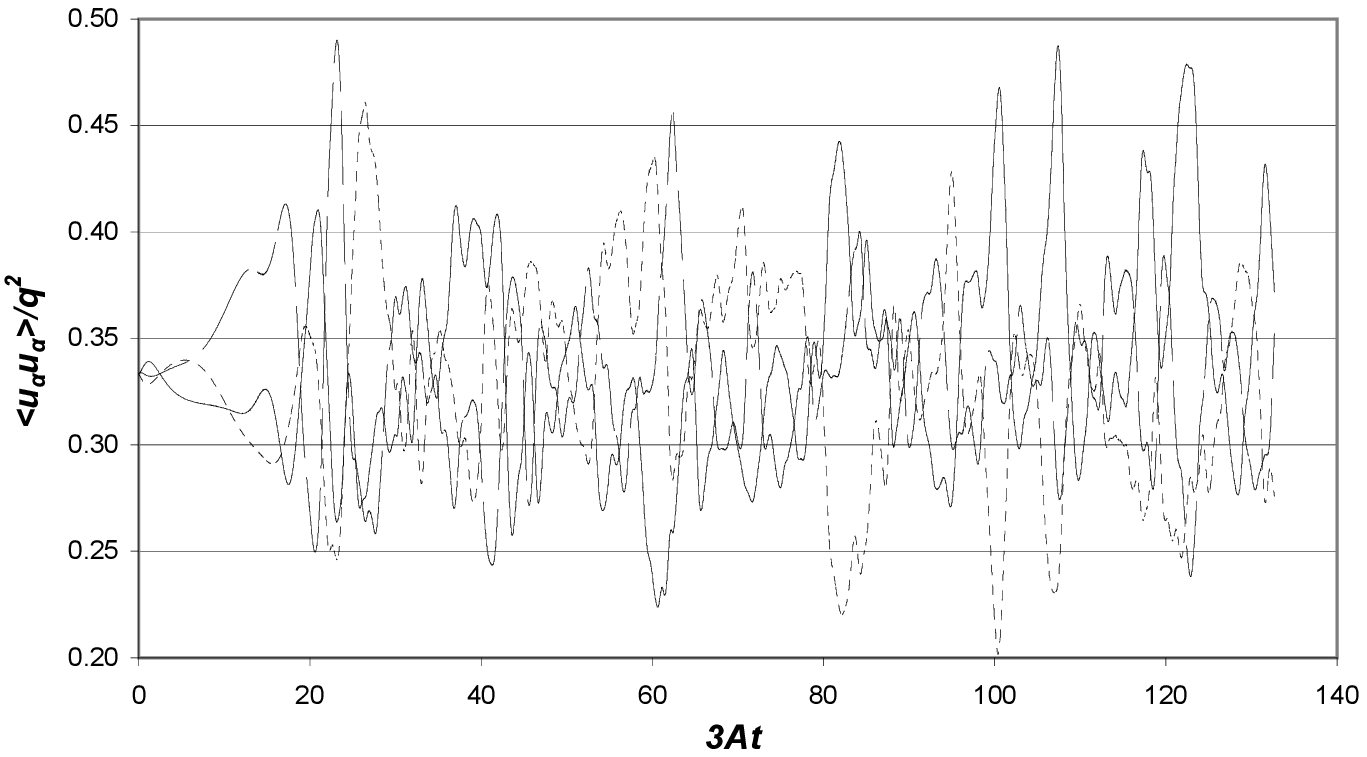}
  \caption{Time evolution of the diagonal components of the normalized Reynolds  stresses, $\langle u_\alpha u_\alpha
  \rangle/q^2$(no sum), from the same run producing the time-series in fig.~\ref{first-figure}.}
  \label{isotropy-figure}
  \end{center}\vspace{-.2in}
\end{figure}

Contemplating the far more complicated motion of the correlators
$\langle X\rangle$ compared to that of $X$, one quickly realizes
that both kinds of correlators obey the same basic dynamical
equations. Thus their difference lies somewhere else. The simplicity
of the motion of $X$ follows from their independence from the
initial conditions of the flow and the symmetries of the dynamical
equations and of the domain. Correlators $\langle X\rangle$ are not
a priori independent of the initial conditions of the flow and none
of the arguments of section \ref{Scaling symmetries and asymptotic
behavior} applies to them. Therefore there are many more and more
complicated solutions $\langle X\rangle$ than $X$.

In particular, in section \ref{Scaling symmetries and asymptotic
behavior} it was emphasized that the symmetries of the cube,
combined with the symmetry (\ref{symmetry}), force a fair amount of
isotropy on the solutions $X$. That will hold only on the average
for the correlators $\langle X\rangle$. This is the actual result of
numerical simulations, fig.~\ref{isotropy-figure}.

We consider the fluctuations of measures of isotropy important for
the following reason. The `fair amount of isotropy' exhibited by $X$
is not so harmless itself: Reasonably, any anisotropy is inherited
by the correlators $\langle X\rangle$ and it is enhanced in their
description of turbulence. The origin of any enhancement of
anisotropy is that in linear forcing there is no intrinsic large
scale at which we feed the flow energy in some isotropic manner; the
large scales are set by the domain itself and the scales comparable
to its size are necessarily not isotropic. Subsequently, anisotropy
is produced and maintained at those and smaller scales through
forcing and cascade. Although the motion of the correlators $X$
suggests that the system cannot be kicked off balance, the produced
anisotropy will cause relatively large fluctuations of all
quantities $\langle X\rangle$ describing turbulence.

In the next section we will try to lend some quantitative support to
this intuitive picture. If deviations from isotropy are related to
the fluctuations of all quantities then any fluctuations should
vanish in a perfectly isotropic setting. That may give us a sense of
what happens when those fluctuations are continually generated. As
we shall see the analysis is interesting in its own right as it
reveals rather important dynamical properties of linearly forced
turbulence.

\section{Self-preserving turbulence and stability of stationarity}
\label{Self-preserving turbulence and linear stability of
stationarity}

Studying the fluctuations around the stationary state is equivalent
to studying the stability of that state as a fixed point of
solutions, in the statistical sense. In fact, this is an alternative
way to look at the main problem we have been concerned with, the
stationary state as an attractor of solutions. Of course, such an
analysis is a very difficult thing to do unless we resort to some
suitable simplification.

According to the plan set at the end of the previous section, we
shall assume that the flow is isotropic. Therefore all correlators
involved, which can be thought of either as ensemble correlators $X$
or box-averages $\langle X \rangle$, are assumed to have the
properties required by that condition. We may then investigate the
fate of any deviations away from the stationary state if the flow
evolves remaining isotropic.

The evolution laws derived in the previous sections can be
alternatively derived in isotropic turbulence from models relating
the kinetic energy $K$ and dissipation rate $\varepsilon$. These
models can be deduced on dimensional grounds, or more systematically
by self-similarity arguments, which are fairly equivalent to the
scaling arguments given here. The latter date back to the work of
von Karman and Howarth~\cite{KH} and Batchelor~\cite{Batchelor48},
see also \cite{Townsend}\cite{Batchelor-Book}. One looks for
self-similar solutions of the equations w.r.t. a single length scale
$L(t)$, `self-preserving' turbulent flows. Assuming that the larger
scales of the flow are evolving in such a self-preserving manner,
 one chooses $L(t)$ to be the integral scale and one obtains a closed
system of equations for the variables $K$ and $\varepsilon$. That
simple model can also be regarded as describing self-preserving
turbulence of all scales but for infinitely high Reynolds numbers,
essentially for inviscid flow.

One can straightforwardly apply the same arguments in the linearly
forced turbulence. The spectral energy balance equation (\ref{energy
balance spectral}) becomes
\begin{equation}\label{energy balance spectral linear}
\partial_t E(k)=-\partial_k  T(k)-2\nu k^2 E(k)+2A E(k)\,.
\end{equation}
The origin of the additional term should be clear. Then the
self-preserving development of the larger scales of the flow
implies, via standard steps which can be found e.g.\ in
\cite{Townsend}\cite{Batchelor-Book}, the model equation
\begin{equation}\label{model eq townsend}
\frac{d\varepsilon}{dt}=-C^A_\varepsilon \frac{\varepsilon^2}{K}+c_1
A \varepsilon\,,
\end{equation}
where $c_1=3$ and $C^A_\varepsilon$ is a dimensionless constant.
Apart from the value of $c_1$, this equation could also have been
guessed on dimensional grounds upon requiring its r.h.s. to be built
out of $\varepsilon$ and $K$ and $A$ and be linear in $A$.

Integrating (\ref{energy balance spectral linear}) over all wave
numbers we obtain again the exact equation (\ref{energy balance
local homogeneity}), which we write down again for convenience,
\begin{equation}\label{energy balance}
\frac{dK}{dt}=-\varepsilon+2AK\,. 
\end{equation}

The system of equations  (\ref{energy balance}) and (\ref{model eq
townsend}) is consistent with a static solution only for
$2C^A_\varepsilon=c_1$. The special case
$C^A_\varepsilon=c_1/2=3/2$, predicted by large scale
self-preservation, implies that $L_\varepsilon=\textrm{constant}$ at
all times the model holds. This is consistent with the general idea
about it. The large scale self-preservation model equation
(\ref{model eq townsend}) and the value $C_\varepsilon^A=3/2$ will
emerge again from a different perspective in section
\ref{discussion}. The model can be easily solved exactly and indeed
predicts that the flow approaches stationarity exponentially fast
for all $C^A_\varepsilon>1$ (the case $C^A_\varepsilon=1$ is
trivially consistent with stationarity).

A more elaborate analysis of the evolution of isotropic turbulence
has been presented in the past in the
Refs.~\cite{George-1987}\cite{George-1989}\cite{Speziale}\cite{george-92:1492}.
In those works the self-similarity hypothesis is applied at the
viscous equations of the flow i.e.\ self-preservation is required to
be true for all scales of turbulence for finite Reynolds. In the
terminology of Ref.~\cite{Speziale}, self-preservation is complete.
An implication of this requirement is that the self-similarity scale
is the Taylor microscale $\lambda_g$.

From the point of view of the linearly forced turbulence all that
sound very relevant and interesting for the following reasons.
First, the linearly forced turbulence comes to the intelligible part
of its course when its Reynolds number approaches the value
(\ref{reynolds at stationarity}) which need not be very high at all;
second, energy is generated uniformly at all points in the domain
and it feels that all scales play
a role in approaching or maintaining stationarity; and third, 
in this problem there is a natural scale for the Taylor length
$\lambda_g$. It is the scale at which energy production balances
dissipation in spectral space, as can be seen by equation
(\ref{energy balance spectral linear}):
\begin{equation}\label{lambda-A}
\lambda_A=\sqrt{\frac{\nu}{A}}\,.
\end{equation}
This is designated as a Taylor microscale because the stationary
state value of the Taylor microscale, $\lambda_{g s}$, is of that
order:
\begin{equation}\label{}
\lambda_{g s}=\sqrt{5}\lambda_A\,.
\end{equation}
This follows from the definition (\ref{taylor lambda general}) of
$\lambda_g$ and the stationary state total balance of energy
production balances dissipation, $2AK=\varepsilon$. For these
reasons the Taylor microscale may be regarded as playing a
particularly significant role in the dynamical aspects of linear
forcing, perhaps quite more significant than in the freely decaying
case. [On the other hand, as everything turns out approaching
constancy, eventually the integral scale might be used as a
self-similarity scale, a choice associated with the model
(\ref{model eq townsend}), providing a more crude and late-time
description of the evolution of the system.] In any case this choice
does provide a closed two-equation model with some interesting
properties.

We may then proceed as follows. There is another exact equation
which we may use along with (\ref{energy balance}). One way to
derive it is to start from the Karman-Howarth equation for linearly
forced isotropic turbulence
\begin{equation}\label{KH-linear-forcing}
\frac{\partial}{\partial t}(q_1^2
f)=\frac{1}{r^4}\frac{\partial}{\partial
r}\Big\{r^4\Big(q_1^3\,h+2\nu\, q_1^2\,\frac{\partial f}{\partial
r}\Big)\Big\}+2A\, q_1^2 f\,,
\end{equation}
applying the definitions (\ref{**}) below. (The spectral energy
balance equation (\ref{energy balance spectral linear}) is a Fourier
transform of (\ref{KH}).) Alternatively, and more instructively, we
can do everything from scratch by differentiating $\varepsilon$
w.r.t. to time using its very definition as an ensemble or
box-average correlator. Then, employing the Navier-Stokes equation
(\ref{1}) and applying the condition of isotropy on any arising
correlator one arrives at
\begin{equation}\label{exact dedt}
\frac{d\varepsilon}{dt}=\frac{7|S|}{3\sqrt{15 \nu}}\,
\varepsilon^{3/2}-\frac{7G}{15}\frac{\varepsilon^2}{K}+2A
\varepsilon\,,
\end{equation}
where $S$ (the velocity gradient distribution skewness) and $G$ are
defined by
\begin{equation}\label{**}
S=\lambda_g^3\left.\frac{\partial^3 h}{\partial r^3}\right|_{r=0}\,,
\qquad G=\lambda_g^4\left.\frac{\partial^4f}{\partial
r^4}\right|_{r=0}\,,
\end{equation}
where $f$ and $h$ are the two-point double and triple point
correlations of the velocity defined in section \ref{Important
quantities and formulas}. Equation (\ref{exact dedt}) can also be
derived by multiplying (\ref{energy balance spectral linear}) by
$2\nu k^2$ and use formulas equivalent to (\ref{**}) and
(\ref{taylor lambda general}) in wave number space.

The system of equations (\ref{energy balance}) and (\ref{exact
dedt}) is not closed, the dependence of $S$ and $G$ on $K$ and
$\varepsilon$ is unknown. Assume now that at some moment $t_0$ the
flow becomes self-similar with a (time-dependent) similarity scale
$\lambda_0$. That means $f$ and $g$ are functions of the
dimensionless coordinate $r/\lambda_0$ alone, modulo a possible
dependence on the initial conditions at $t_0$. Now (\ref{taylor
lambda general}) tells us that $\lambda_0/\lambda_g$ must be a
\emph{constant}, depending only on the initial conditions at $t_0$.
Thus the similarity scale is indeed the Taylor microscale. Then by
(\ref{**}) we have that $S$ and $G$ are constant and equal to the
values they have at $t_0$: $S=S_0$ and $G=G_0$. Now the system
(\ref{energy balance}) and (\ref{exact dedt}) is closed and we may
study it.

Let us denote the stationary state values of the dissipation rate
and kinetic energy by $\varepsilon_s$ and $K_s$. Of course they are
related by $\varepsilon_s=2A K_s$. We would like to study the
stability properties of $\varepsilon=\varepsilon_s$ and $K=K_s$ as a
complete self-preserving solution of the system  of equations
(\ref{energy balance}) and (\ref{exact dedt}).

It will be convenient to define the quantity
\begin{equation}\label{}
g \equiv \frac{7G_0}{15}\,.
\end{equation}

First of all, equation (\ref{exact dedt}) implies that
\begin{equation}\label{}
\varepsilon_s^{1/2}=2A\frac{3\sqrt{15\nu}}{7|S_0|}\,(g-1)\,,
\end{equation}
which implies that
\begin{equation*}\label{}
g>1\,.
\end{equation*}
It is useful to relate the value of $g$ to the Taylor-scale Reynolds
number
$\textrm{Re}_\lambda=(\frac{20}{3}\textrm{Re}_L)^{\frac{1}{2}}$. By
(\ref{Le}) we find that its stationary value $\textrm{Re}_{\lambda
s}$ reads
\begin{equation}\label{Re-lambda-s}
\textrm{Re}_{\lambda s}=\frac{30}{7|S_0|}\, (g-1)\,.
\end{equation}

Define now small fluctuations $\xi$ and $\zeta$ of $\varepsilon$ and
$K$ around their stationary values:
\begin{equation}\label{}
\varepsilon=\varepsilon_s\,(1+\xi)\,, \quad K=K_s\,(1+\zeta)\,.
\end{equation}
Inserting these expressions into (\ref{energy balance}) and
(\ref{exact dedt}) and keeping only linear terms we obtain the
following system of equations:
\begin{align}\label{linear stability system}
& \frac{d\xi}{dt}=-A(1+g)\, \xi+2Ag\, \zeta\,, \\
& \frac{d\zeta}{dt}=-2A\, \xi+2A\, \zeta\,. \nonumber
\end{align}
Its eigenvalues $\Gamma$ read
\begin{equation}\label{gamma}
\Gamma=\frac{1}{2}A\left(-(g-1)\pm \sqrt{(g-1)(g-9)}\right)\,.
\end{equation}
By $g>1$ we see that the real part of both eigenvalues is always
negative. Fluctuations around the stationary state die out
exponentially fast. That is, \emph{modulo finite domain effects, the
stationary state is stable as a complete self-preserving isotropic
solution}. We may also view this result as providing further
evidence that the stationary state is the natural final state of the
linearly forced turbulence. [Presumably, one may observe that
exponentially fast approach to the stationary state is also the
prediction of the simpler model (\ref{model eq townsend}).]

The previous analysis can be alternatively understood as follows. In
order to derive the previous results we have assumed perfect
isotropy. A reasonable assumption about the deviations from isotropy
is that they originate from scales of order $l$. That means,
according to our conclusions in the previous section, that the same
can be said about the fluctuations around the stationary state. That
is, one may attribute the generation of fluctuations to the
interaction of the larger eddies with the periodicity i.e.\ the
restriction to their size. Then, through both forcing and cascade,
fluctuations are generated at all scales from $l$ down to a certain
scale where isotropy becomes a good approximation. There things are
different. We may define correlators as spatial averages $\langle X
\rangle_V$ over volumes $V$ smaller than that maximum isotropic
scale i.e. within these volumes turbulence is isotropic (meaning
homogeneity as well) to a good approximation. Then $K$ and
$\varepsilon$ understood as spatial averages $\langle X \rangle_V$
obey similar equations to those studied above. The entire previous
analysis goes through. That finally means that at adequately small
scales the fluctuations are strongly suppressed, but at all higher
scales are maintained through forcing and cascade. The maximum
isotropic scale should be (very) roughly related to the
characteristic Taylor microscale of linear forcing
$\lambda_A=(\nu/A)^{\frac{1}{2}}$, as below that scale dissipation
becomes stronger to energy production.

We may investigate the linear system (\ref{linear stability system})
a bit further. Though this system meant to serve us mainly for
qualitative considerations, regarding the stability of the constant
solution $K=K_s$ and $\varepsilon=\varepsilon_s$,  there are some
amusing remarks to be made about it solutions on the quantitative
side. In the range $1<g<9$ the eigenvalues $\Gamma$ are complex
numbers. If we take for definiteness $|S_0|= 0.5$, that means that
when $\textrm{Re}_\lambda<69$ the fluctuations are damped
oscillations. [Presumably, this emergence of oscillations is a
qualitative difference between the complete self-preservation model
and the simpler model (\ref{model eq townsend}).] Inserting the
solutions
$\zeta=\zeta_0 e^{\Gamma t}$
and $\xi=\xi_0 e^{\Gamma t}$, 
for positive frequency into any of the equations (\ref{linear
stability system}) we obtain the phase difference and the relative
amplitude of $\varepsilon$ and $K$:
\begin{equation}\label{fluctuations ratio}
\xi_0=\sqrt{g}\, e^{-i \phi}\, \zeta_0\,,
\end{equation}
where $\phi$ is given by
\begin{equation}\label{tanphi}
\tan \phi=\frac{\sqrt{(g-1)(9-g)}}{g+3}\,.
\end{equation}
As expected the dissipation $\varepsilon$ evolves with a phase delay
w.r.t. the kinetic energy $K$ and the energy production $2AK$. This
corresponds to a time-delay $\phi/|\textrm{Im}\Gamma|$. The period
of these damped oscillations is of course
$2\pi/|\textrm{Im}\Gamma|$.

\begin{figure}[t]\vspace{.0in}
\begin{center}
  \includegraphics[height=0.265\textwidth,
  angle=0]{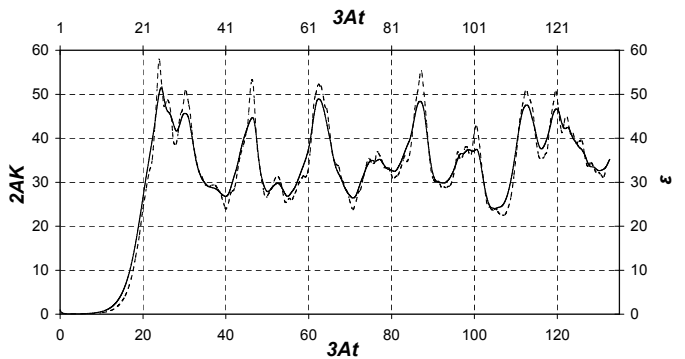}\hspace{+1in}
  \\
  \caption{The time evolution of the dissipation rate $\varepsilon$ which is shown in
  fig.~\ref{first-figure}, is
   shifted in this figure by one unit of dimensionless time $3At$.}
  \label{second-figure}
  \end{center}\vspace{-.20in}
\end{figure}

In fig.~\ref{first-figure} we plotted the energy production $2AK$
and dissipation rate $\varepsilon$ against time in units of
$(3A)^{-1}$. Now let us shift the evolution of dissipation by one
unit of time to off-set its delay. The result is given in the
fig.~\ref{second-figure}. One observes that after that shift the
complicated oscillations appear in phase to a considerable degree of
accuracy. Curiously, the time-delay $\phi/|\textrm{Im}\Gamma|$ in
units of $(3A)^{-1}$ decreases from $1.5$ to $0.5$ in the range
$1<g<9$. Also the period $2\pi/|\textrm{Im}\Gamma|$ is roughly an
order of magnitude higher than $(3A)^{-1}$ for most values of $g$,
which as a number is not in disagreement with the picture in
fig.~\ref{second-figure}. Given that these numbers derive from a
model which does not interact with the source of the fluctuations,
it seems interesting that the oscillations it implies may
encapsulate certain features of the actual fluctuation. There
certainly is no identification between the actual fluctuations and
those oscillations. For example, when $g \sim 9$ that is
$\textrm{Re}_{\lambda s} \sim 70$ the damped oscillations are
replaced by a purely decaying exponential, a qualitative change in
the behavior which cannot be traced in the DNS results of the
Refs.~\cite{Rosales}\cite{Akylas1}\cite{TurbulenceIII}.


\begin{figure}[t]\vspace{-.0in}%
\begin{center}
  \subfloat[]{\includegraphics[height=0.23\textwidth,  angle=0,%
  clip=]{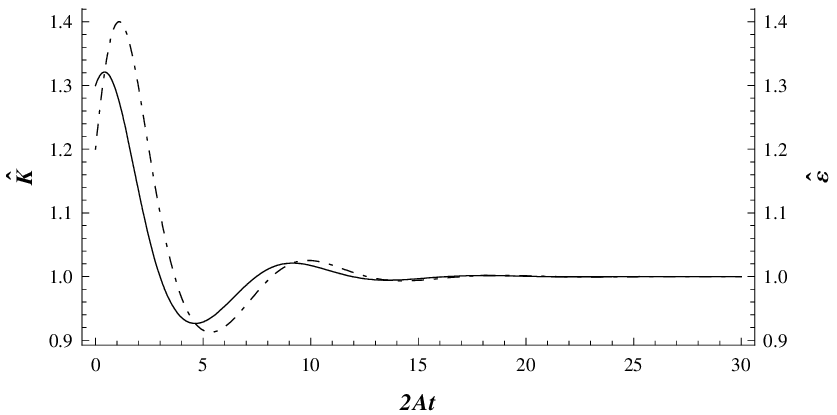}}
 \\[4pt]%
  \subfloat[]{\includegraphics[height=0.23\textwidth, angle=0,%
  clip=]{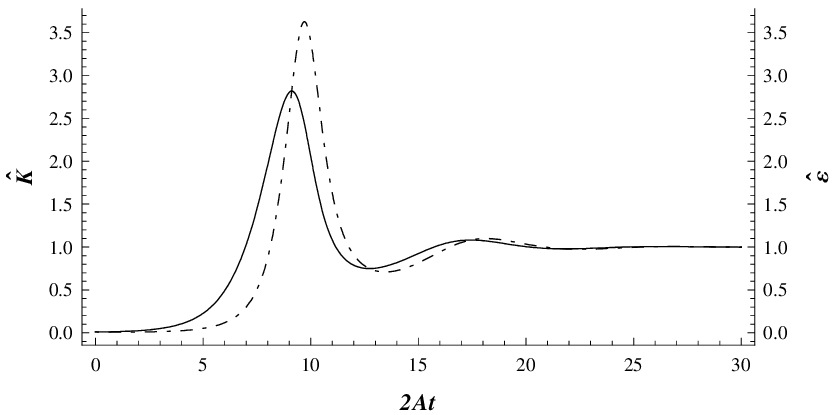}}
\\[4pt]%
  \subfloat[]{\includegraphics[height=0.23\textwidth, angle=0,%
  clip=]{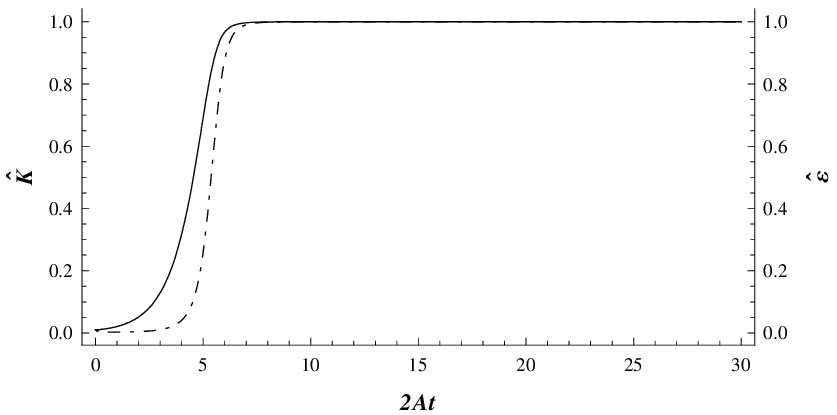}}
  \end{center}
  \vspace{-0.12in}%
  \caption{Time evolution of the kinetic energy (solid line)
  and dissipation (dashed line) normalized by their stationary state
  values from numerical solutions of the system
  (\ref{non-linear}).
   Figure (a): initial conditions $\hat K(0)=1.3$ and $\hat \varepsilon(0)=1.2$ for $g=2.2$ that is $\textrm{Re}_{\lambda s} \sim 10$.
  Figure (b): $\hat K(0)=0.01$ and $\hat \varepsilon(0)=0.01$ for the same value of $g$. Figure (c): $\hat K(0)=0.01$ and $\hat
  \varepsilon(0)=0.01$ for $g=9.9$ that is $\textrm{Re}_{\lambda s} \sim 85$.}%
  \vspace{-.15in}%
  \label{K-e-num}
\end{figure}



The previous remarks derive from the quantitative characteristics of
small fluctuations, and we may have over-extended the applicability
of the related formulas. Arbitrary fluctuations are described by the
solutions of the full non-linear model (\ref{energy balance}) and
(\ref{exact dedt}). This needs to be solved numerically. In terms of
the dimensionless (hatted) kinetic energy, dissipation rate and
time, defined respectively by $K=K_s\, \hat K$,
$\varepsilon=\varepsilon_s\, \hat \varepsilon$ and $\hat t=2A t$,
the non-linear model reads
\begin{align}\label{non-linear}
& \frac{d \hat K}{d\hat t}=-\hat \varepsilon+\hat K\,, \\ & \frac{d
\hat \varepsilon}{d\hat t}=(g-1)\, \hat \varepsilon^{3/2}-g
\frac{\hat \varepsilon^2}{\hat K}+\hat \varepsilon\,. \nonumber
\end{align}
The parameter $g$ is related to $\textrm{Re}_{\lambda s}$ by
(\ref{Re-lambda-s}) and we again take for definiteness $|S_0|=0.5$.

The system (\ref{non-linear}) is solved using the software
\emph{Mathematica}. We consider a few specific cases. First, the
difference of the initial conditions from the stationary state
values is such that to imitate the size of the observed
fluctuations. This is shown in fig.~\ref{K-e-num}a. Second, the
kinetic energy $\hat K$ and dissipation rate $\hat \varepsilon$
start off from very close to zero, shown in fig.~\ref{K-e-num}b. For
those two cases we have chosen an adequately small Reynolds number
so that oscillations to be visible. Finally we consider the effect
of higher Reynolds numbers. An evolution of $\hat K$ and $\hat
\varepsilon$ for $\textrm{Re}_{\lambda s} \sim 85$ is shown in
fig.~\ref{K-e-num}c.

The result is that the picture is not qualitatively different than
the one obtained from the small fluctuations. In the
figs.~\ref{K-e-num}a and \ref{K-e-num}b, the time-delay of the
dissipation relatively to the kinetic energy and the period of the
damped appear essentially as predicted previously, and the
oscillations of the dissipation are consistently larger as implied
by equation (\ref{fluctuations ratio}). On the other hand
fig.~\ref{K-e-num}b shows a particular behavior of the non-linear
solutions: if the initial condition is far away from the stationary
state values the system undergoes large fluctuations before settling
to those values. The fig.~\ref{K-e-num}c shows that increasing the
Reynolds number any wiggling of the curves due to oscillatory
behavior diminishes to extinction, which is again what we expected.

\section{discussion}
\label{discussion}

Direct numerical simulations of turbulence forced by the linear
forcing scheme exhibit a not entirely expected stationary late-time
state. The stationary phase is essentially quasi-stationary: all
quantities have relatively large fluctuations, though their
time-average can be predicted fairly well. In the present work we
have attempted to understand how these phenomena are rooted in the
properties of the system. This was done by using the symmetries of
the dynamical equations of the problem, as well as the symmetries
associated with the boundary conditions i.e.\ the size and the
symmetries of the cubic domain; also, using special dynamical
properties of the system derived under usually employed conditions
such as exact isotropy or self-similarity. In this problem there are
few and specific scales: the domain size $l$, the constant rate $A$
and the viscosity $\nu$. Out of them derive a Taylor microscale
$\lambda_A$ and a Reynolds number $\textrm{Re}_A$. These quantities
control the major (intelligible) features of linearly force
turbulence evolution.

The importance of the finiteness of the domain and its effects
cannot be over-emphasized in the linearly forced turbulence. In a
limited bandwidth forcing scheme, deterministic or stochastic, the
inverse wave numbers at which one forces the flow imitate, very
roughly, the scale of a physical stirring of an incompressible fluid
existing in slightly larger 'box'. In linear forcing there is no
such intrinsic scale. This simplifies things in some sense because
there is no interaction between the forcing and domain size scales.
On the other hand it is left entirely to the domain to set the large
scales, becoming an essential part of the forcing itself. Also the
large scale is introduced geometrically as a matter of size and not
dynamically as in the bandwidth schemes, and there is no actual
control over the extent forcing is consistent with isotropy.
Turbulence is expected to behave quite differently under linear
forcing than under a bandwidth scheme. There some additional
interesting properties of linear forcing we have not yet commented
on.  These properties can be associated with the effects of the
finite domain size, and also show an at least formal affinity of the
linear forcing to freely decaying turbulence, than to the bandwidth
forcing schemes.

Denote by $\Delta u_l$ the longitudinal velocity difference. The
second and third order structure functions are related to the
correlation functions $f$ and $h$, introduced in section
\ref{Homogeneous and isotropic turbulence}, by $\overline{(\Delta
u_l)^2}=2q_1^2(1-f)$ and $\overline{(\Delta u_l)^3}=6 q_1^3 h$. For
adequately high Reynolds numbers there is a range of distances (the
inertial range) where $\overline{(\Delta u_l)^2}=C_2 (\varepsilon
r)^{2/3}$, where $C_2$ a constant. Consider first decaying
turbulence. It evolves according to the power laws (\ref{L K of
gamma}), the integral scale is proportional to $t^\gamma$.  The law
for $\varepsilon$ can be deduced. It is then straightforward to show
that they satisfy the $K-\varepsilon$ model equation (\ref{model eq
townsend}) for
\begin{equation}\label{ce}
C_\varepsilon=\frac{3-2\gamma}{2-2\gamma}\,,
\end{equation}
and of course $A=0$. Using the Karman-Howarth equation (\ref{KH}) it
then straightforward to show~\cite{Lindborg}\cite{lund-2} that for
\emph{very high but finite} Reynolds numbers, and within the
inertial range (more specifically as long as $r/\lambda_g$ is a
number of $O(1)$), the two-thirds law of the second order structure
function implies specific finite Reynolds number corrections to the
four-fifths law of the third order structure function, of
$O(\textrm{Re}_\lambda^{-2/3}$). The result
is~\cite{Lindborg}\cite{lund-2}
\begin{align}\label{4/5-corrections}
& \qquad \overline{(\Delta u_l)^3}= -\frac{4}{5}\varepsilon r \times \\
& \left(\!1-\frac{5\times 15^{\frac{2}{3}}}{17}C_\varepsilon C_2
\textrm{Re}_\lambda^{-\frac{2}{3}}
\Big(\!{\frac{r}{\lambda_g}}\!\Big)^{\frac{2}{3}}-\Big(\!\frac{25}{3}\!\Big)^{\frac{1}{3}}C_2
\textrm{Re}_\lambda^{-\frac{2}{3}}\Big(\!\frac{r}{\lambda_g}\!\!\Big)^{-\frac{4}{3}}
\right). \nonumber
\end{align}

Consider the same question in the linearly forced turbulence. One
may follow the same steps, starting from the Karman-Howarth equation
with linear forcing, equation (\ref{KH-linear-forcing}). One finds a
result entirely similar to (\ref{4/5-corrections}) upon replacing
\begin{equation}\label{sub}
C_\varepsilon \to
-\frac{K}{\varepsilon^2}\frac{d\varepsilon}{dt}+\frac{3AK}{\varepsilon}\,.
\end{equation}
Observe now that if we think of the r.h.s. of this substitution as a
constant, then we re-discover the model equation (\ref{model eq
townsend}); the constant is what we denoted there by
$C_\varepsilon^A$. Equation (\ref{model eq townsend}) is derived
assuming self-similarity (self-preservation) of the larger scales of
turbulence with respect to the integral scale $L$ for high Reynolds
numbers, in both the linearly forced ($A \neq 0$) and freely
decaying case ($A=0$). In all, by self-preservation we obtain a
similar result of the form (\ref{4/5-corrections}) in both kinds of
turbulence, differing only in the value of the constants
$C^A_\varepsilon$ and $C_\varepsilon$. On the linearly forced side,
self-preservation requires $C_\varepsilon^A=3/2$ and equation
(\ref{model eq townsend}) and (\ref{energy balance}) require that
$L=\textrm{constant}$. At first sight there is no such restriction
on the freely decaying side. In all, there appears to be a
correspondence between linearly forced and freely decaying
turbulence, though this correspondence appears inexact.


Now if we require $C_\varepsilon^A=C_\varepsilon$ then by (\ref{ce})
we have that $\gamma=0$. In other words, if the decaying turbulence
evolves according to $L \sim \textrm{constant}$ (and $K \sim
t^{-2}$) then its structure function expression
(\ref{4/5-corrections}) is exactly similar to that of the linearly
forced turbulence. That is, the correspondence between the two flows
can be exact.

The $K \sim t^{-2}$ evolution is too fast compared to the usually
observed decay laws, discussed in section \ref{Scaling symmetries
and power laws}. Such power laws can be reproduced if choose the
constant $C_\varepsilon$ to be different than 3/2, a fact regarded
as an imperfection of the correspondence in the Ref.~\cite{Lund},
where it was first pointed out. On the other hand the origin and the
nature of the correspondence seem to have been overlooked in
\cite{Lund}.

The key role is played again by the finiteness of the domain. As
emphasized in section \ref{Effects of the finite domain} a container
is a necessary thing when turbulence is linearly forced. Lacking an
intrinsic length scale, linear forcing essentially requires a large
scale to be provided by the boundary conditions. It is therefore not
much of a surprise that similarities between linearly forced and
freely decaying turbulence are more detailed when the decaying side
evolves in a way consistent with the existence of a container: For
adequately high Reynolds numbers that means $L \sim
\textrm{constant}$ (and the rest of the (\ref{L K of
gamma})-(\ref{Re of gamma}) power laws for $\gamma=0$). Then the
mathematics of self-similarity of turbulence with respect to the
scale $L$ imply exactly the same formula (\ref{4/5-corrections}) for
both kinds of turbulence.

The next obvious question is, what kind of modifications does linear
forcing need in order to reproduce aspects of a generic decaying
turbulence, associated with (\ref{ce}) and an evolution law $L \sim
t^\gamma$? Two immediate guesses are to consider a time-dependent
rate $A=A(t)$ or, to consider a time-dependent box whose size $l$
evolves according to $l \sim t^\gamma$. The analysis of such
possibilities is left for future work.






\bibliography{stationarity}      
\bibliographystyle{utcaps}
\end{document}